# A modification of the $h$-index:

# the $h_m$-index accounts for multi-authored manuscripts

Michael Schreiber

Institut für Physik, Technische Universität Chemnitz, 09107 Chemnitz, Germany

Tel.: (+49) 371 531 21910

Fax: (+49) 371 531 21919

e-mail: schreiber@physik.tu-chemnitz.de

**Abstract**

In order to take multiple co-authorship appropriately into account, a straightforward modification of the Hirsch index was recently proposed. Fractionalised counting of the papers yields an appropriate measure which is called the $h_m$-index. The effect of this procedure is compared in the present work with other variants of the $h$-index and found to be superior to the fractionalised counting of citations and to the normalization of the $h$-index with the average number of authors in the $h$-core. Three fictitious examples for model cases and one empirical case are analysed.

## 1. Introduction

The $h$-index has been designed by Hirsch (2005) to measure the impact of a scientist's publications in terms of the citations received. It is defined as the highest number of papers of a researcher that have been cited *h* or more times. In the last two years it has been analysed for various groups in different fields and accumulated more than 70 citations already, thus enhancing Hirsch's Hirsch-index. Among the possible disadvantages of ranking scientists in terms of their *h*-index it has often been mentioned that the *h*-index does not take into account multiple co-authorship (Batista et al. 2006, Bornmann and Daniel 2007, Burrell 2007, Hirsch 2005, Hirsch 2007, Imperial and Rodriguez-Navarro 2007). Already Hirsch (2005) proposed "to normalize *h* by a factor that reflects the average number of co-authors". This idea has been applied by Batista et al. (2006) dividing *h* by the mean number of authors of the papers in the *h*-core, i.e., in the *h*-defining set of papers. The resulting so-called $h_I$-index has been determined for a large community of Brazilian researchers (Batista et al. 2005). In that study the authors have already cautioned against the problem that the average is sensitive to extreme values and therefore the normalization with the mean number of authors disfavours people with some papers with a large number of co-authors. For the same reason, the effect of the other "extreme" is that the influence of single-author publications to one's *h*-index can be rather strongly reduced. Moreover, as demonstrated below, a peculiar behaviour can be observed in certain cases, because the $h_I$-index might decrease when a paper with many authors advances into the *h*-core by attracting additional citations.

One way to overcome this difficulty would be to count the citations fractionally, i.e., to divide the number of citations by the number of authors for each paper. One could then define a Hirsch-type index $h_f$ as that number of papers for which this ratio is at least equal to $h_f$. However, this has the disadvantage that for a determination of $h_f$ the publications have to be rearranged into a new order according to this quotient. It also leads to the strange effect that highly cited papers may not contribute to the index because they have a large number of authors, so that they drop out of the core by the rearrangement.

Burrell (2007) noted that "if any of an author's papers contributing to the $h$-index is multi-authored, then applying any sort of discounting could well remove that paper from the $h$-core". I will demonstrate in this manuscript that there is a way of discounting which avoids this problem.

In fact, all these problems to not occur, when a fractionalised counting of the papers is utilized to determine the index. Adding the fractional counts yields a reduced number which can be interpreted as a reduced rank or an effective rank. It is then straightforward to define a Hirsch-type index which I label $h_m$-index (because it accounts for multiple authorship) as that reduced number of papers that have been cited $h_m$ or more times (while the other papers have been cited not more than $h_m$ times).

I have recently analysed the $h_m$-index of the citation records for 8 prominent physicists (Schreiber 2008a), resulting in a different ranking than the original $h$-indices. For the present manuscript, three fictitious examples for model cases have been constructed. The first example is utilized to demonstrate the determination of the $h_m$-index and the other variants of the Hirsch index mentioned above. The other two cases reflect somewhat exceptional data sets which I have constructed to show extreme situations in order to point out unusual or strange behaviour which can occur in the calculation of the $h_f$-index and the $h_I$-index. This allows me to argue that the modified index is superior to the other variants, when one wants to take multiple authorship into account. For visualization purposes I have also included the analysis of my own citation record as an empirical example. I further compare the modified index $h_m$ with the $h_f$-index of fractionalised counting of citations, which was not discussed in my previous analysis (Schreiber 2008a).

During the reviewing process of the present manuscript, I became aware of another paper (Egghe 2008) in which a mathematical theory of several variants of the Hirsch index is presented in case of fractionally counting, including the here discussed $h_\mathrm{m}$-index and $h_\mathrm{f}$-index. In that paper also some fictitious examples and one empirical case are analysed. Long before the Hirsch index was proposed other methods of treating the multiple-author problem have been discussed (Egghe et al. 2000) showing that one particular method does not contain an absolute truth and that therefore it is unclear which distribution of the credit to co-authors is the correct distribution. In the present manuscript I argue that at least for the Hirsch index the proposed modification is more appropriate than the other variants.

**2. Three fictitious examples for model cases**

In order to demonstrate the determination of the $h_\mathrm{m}$-index and its behaviour in contrast to the $h_\mathrm{I}$-index and the $h_\mathrm{f}$-index, let us consider as an example a simple data set with 8 publications as listed in table 1, where the papers are ranked according to their number of citations. Obviously $h = 5$. The average number of authors of these papers is $12/5 = 2.4$, so that $h_\mathrm{I} = 5/2.4 = 2.08$, which is rather small and disregards the sixth and seventh paper with citation counts of $c = 3$, i.e. with citation counts larger than $h_\mathrm{I}$. Counting the citations fractionally as in the fourth column of table 1, one notes that the fourth paper dropped out of the core so that one obtains the $h_\mathrm{f}$-index after rearranging the papers according to the values in the fourth column as $h_\mathrm{f} = 4$.

For the determination of the $h_\mathrm{m}$-index the papers in table 1 have to be counted fractionally according to (the inverse of) the number of authors. This yields the effective rank $r_\mathrm{eff}$ given in the last column of table 1. Going down the table, one can see that in the seventh row this effective rank has reached the number of citations, i.e., in this row the criterion for the $h_\mathrm{m}$-index is fulfilled, consequently $h_\mathrm{m} = 3$. This means that due to the fractionalised counting of publications, two more papers have entered into the $h_\mathrm{m}$-core.

In order to demonstrate the differences of the indices more clearly, I have constructed another example of a data set which is rather extreme. It consists of 3 publications with 9 citations and 3 authors each, 4 publications with 8 citations and 2 authors each, and 4 single-author papers with 7 citations. These papers are ranked according to their number of citations in table 2 and it is obvious that $h = 7$. There are ambiguities in the ranking of the papers in table 2, because publications with the same number of citations cannot be distinguished. This ambiguity is inherent already in the original definition of the $h$-index (Hirsch 2005). A reasonable solution to this problem would be to sort these papers by publication date, e.g. placing the most recent publication first, because it has attracted a higher number of citations per year since it was published.

The average number of authors of the first 7 papers in table 2 is $17/7 = 2.43$ which yields $h_I = 7/2.43 = 2.88$, which is a very small value especially in view of the fact that there are 4 further papers in the data set with significantly more citations than 2.88. This appears to me to be inappropriate.

Counting the citations fractionally as in the fourth column of table 2, one immediately sees that in this special case the ranking is reversed and the last 4 papers in table 2 yield the value $h_f = 4$. Now the highly cited papers do not contribute to the index at all. In my opinion this is also not appropriate. Moreover, in a more realistic data set the $h_f$-index could not be read off the table easily unless the rows would first be rearranged according to the values in the fourth column. By the way, this might lead to an enhancement of the $h_f$-index, if there were further single-author papers with 5,6, or 7 citations beyond the first 11 papers ranked in table 2.

Counting the papers in table 2 fractionally yields the effective rank $r_{eff}$ as given in the last column of the table. In this extreme case the effective rank reaches the number of citations in the last row, so that $h_m = 7$. All 11 papers contribute to this $h_m$-index, but further papers beyond $r = 11$ would not contribute.

Let us consider another constructed example of a rather extreme model data set which consists of 4 papers with 16 citations and two authors each, 4 single-author publications with 8 citations, and one further paper with 6 citations and 5 authors, see table 3. In this case the first 8 papers contribute to the $h$-, to the $h_f$-, as to well as the $h_m$-index, and the ninth paper and any further papers beyond the rank $r = 9$ do not contribute so that $h = 8$, $h_f = 8$, and $h_m = 6$. The average number of authors of the first 8 papers is 1.5, which yields $h_I = 5.33$.

An interesting behaviour can be observed, when the citation count of the ninth paper of this data set is increased, as shown in table 4. Already for $c = 7$ it contributes to the $h_m$-index, although the contribution is small due to the large number of authors. Increasing the citation count to $c = 8$ does not demand a change of any of the indices, because the ranking in table 3 need not be changed. However, there is an ambiguity, because as mentioned above the definition of the $h$-index does not specify how to arrange papers with an equal number of citations. Therefore the ranking *can* be changed, and the ninth paper might be advanced to rank 5. This ambiguity has no effect on the $h$-index, because advancing this paper to fifth position pushes the previously eighth paper out of the $h$-core. However, this rearrangement does have an effect on the average number of authors of the papers in the $h$-core which increases to $16/8 = 2$, and as a consequence the $h_I$-index changes dramatically: it drops to $h_I = 4$, see table 4. When the citation count is further increased, there is no more ambiguity: the previously ninth paper has now definitely advanced to the fifth rank and the $h_I$-index has definitely dropped to the lower value. This means that increasing the number of citations has led to a decrease of an index which is supposed to measure the impact of the publications in terms of the number of citations. In my opinion this is a very strange behaviour indeed.

In order to contribute to the $h_f$-index it would be necessary that this paper received at least 40 citations as indicated in the last row of table 4. But even then it would not enhance the index, but only push one of the other papers out of the $h_f$-core.

Of course the ambiguity also influences the effective rank. When the original ranking is changed and the ninth paper advances to rank 5, the corresponding rank of this paper changes from 6.2 to 2.2, compare table 4. However, this does not influence the $h_m$-index, because as before all nine papers contribute to the $h_m$-index. So the strange behaviour found for the $h_I$-index is not found here. When the paper finally advances to the first position its effective rank is of course 0.2, but again there is no influence on the $h_m$-index.

That such a large increase in the number of citations has no influence on the index may seem inappropriate. But this is a problem already inherent in the original definition of the $h$-index, as indicated in the fourth column of table 4. Likewise the $h_f$-index does not change, see the sixth column of table 4. A straightforward way to take the number of citations of the highly cited papers into account is provided by means of the $g$-index (Egghe 2006) as discussed and compared to other variants of the $h$-index by Jin et al. (2007) and Schreiber (2008b).

**3. An empirical example**

To demonstrate the determination of the $h_m$-index for an empirical data set I have analysed my own citation record obtained from the general search from the Science Citation Index provided by Thomson Scientific in the ISI Web of Science excluding homographs. The results are displayed in the upper histogram in figure 1, and the intersection with the white line yields the value $h = 28$ for my Hirsch index. It is easy to visualize the effect of the fractionalised counting of the papers because this yields just narrower bars for multi-author publications. As a result the histogram is substantially compressed towards the left and the reduced numbers of the 28 papers in the $h$-core yield only $r_{eff} = 11.53$, as indicated in figure 1. Of course, for this effective rank the number of citations is well above the white line in figure 1. Consequently, several papers with lower citation counts have to be taken into account for the $h_m$-core in the same way as this happened in the constructed data sets in tables 1 and 2. In my case the resulting $h_m$-core comprises 43 publications. The reduced numbers add up to an effective rank $h_m = 18.48$, which can be derived from the intersection of the middle histogram in figure 1 with the white line. Of course, the number of 43 publications in the $h_m$-core means that a significantly larger number of papers in the Web-of-Science data base has to be checked for homographs. This so-called precision problem is quite severe in my own case where more than two thirds of the papers found in the Web of Science with citation counts larger than $h_m = 18$ have not been (co-)authored by myself but by scientists with the same surname and the same initial.

The graphical approach that was used for the determination of the $h_m$-index can also be utilized to visualize the derivation of the $h_I$-index. Again counting each paper fractionally but now according to the mean number 2.89 of authors in the $h$-core, and simultaneously scaling the number of citations by the same mean number yields the lower histogram in figure 1 and now its intersection with the white line reflects the $h_I$-index. In my opinion this derivation shows that the normalization of the $h$-index by the mean number of authors leads to an unreasonably strong reduction of the index, because it effectively means that not only the citations are fractionally counted, but also the papers are fractionally counted at the same time. As a result, $h_I = 9.69$ is so small that a large number of publications with larger citation counts exist in the data set, but are not considered because they have received less than $h$ citations. The same problem occurred in all three example cases above, compare tables 1, 2 and 3. In my own case there are 64 publications with citation counts between $h_I$ and $h$, and it is inappropriate that not even the four single-author papers among them nor the five two-author publications with citation counts between $2*h_I$ and $h$ have an effect on the $h_I$-index. These extreme effects are mitigated when the square root of the mean number of authors of the papers in the $h$-core is utilized for the normalization of the $h$-index. This corresponds to the $h_P$-index which was proposed for evaluating the "pure" contribution of a given author (Wan et al. 2007) if the papers are counted fractionally. However, the problems mentioned for the $h_I$-index remain, although in weaker form.

One can observe in figure 1 that the horizontal compression of the upper histogram to the middle histogram is not as strong as that to the lower histogram. This reflects the above statements that the mean is sensitive to extreme values and thus the compression by the average number of authors is stronger than the average compression by the number of authors.

In order to determine the $h_f$-index, one has to divide the number of citations for each paper by the corresponding number of authors. The result is shown in figure 2. As demonstrated already for the example of the fictitious model cases in tables 1 and 2, this division can necessitate a substantial rearrangement, when the papers are ranked according to the quotient. This is visualized in figure 2 where the intersection of the white line with the lower histogram now yields the value $h_f = 18$. The largest original rank of a paper that finally ended up in the $h_f$-core after the rearrangement was $r = 44$. Consequently the precision problem is also much more severe for the calculation of the $h_f$-index than for the determination of the $h$-index, because a much larger number of papers have to be checked to exclude homographs. In principle, all 49 papers with at least 18 citations have to be considered, but the difficulty is that this value is known only after the rearrangement. Accordingly the precision problem increases substantially also compared with the precision problem for the calculation of the $h_m$- index, and what is worse, often the checking will turn out to be unnecessary for part of the considered range a posteriori, namely after the rearrangement.

**4. Further discussion and summary**

In the present investigation I have compared different ways to take the number of authors into account in the determination of Hirsch-type indices. I have demonstrated that the normalization of the *h*-index by the mean number of authors of the publications in the *h*-core leads to an excessive reduction of the index, because it effectively means a fractionalised counting of the citations as well as a fractionalised counting of the publications. Counting fractionally only the number of citations seems to be an appropriate approach, but it is rather impractical, because of the required rearrangement of the papers which also means that a large number of papers have to be checked for homographs, even though these papers do not end up in the $h_f$-core after the rearrangement.

Therefore I have proposed (Schreiber 2008a) to fractionalise the number of publications by counting each paper only according to the (inverse of the) number of authors. Then the order of the papers is not changed, and the effective number of publications allows a straightforward determination of the respective index, which I have labelled $h_m$-index. The precision problem is somewhat enhanced, as additional papers enter into the $h_m$-core in comparison with the *h*-core, but all papers up to the rank *r* for which $r_{eff}(r) = h_m$ contribute, so that at least no unnecessary checking for homographs is required.

One further advantage of the $h_m$-index is worth noting: it allows for a straightforward aggregation of data sets, which is useful, e.g., if one wants to quantify the combined index of several people like all scientists in an institute. For example, if two researchers of that institute have written a paper together, it contributes to their $h_m$-indices exactly two times one half, as it should (provided that its citation count is large enough). On the other hand it would be fully taken into account twice for the $h$-indices, which appears to be unreasonably high. For the $h_I$-indices the count would depend on the average number of co-authors of either scientist or, in the worst case, on the average number of co-authors of all people in the institute, which in my opinion is completely unjustified.

In conclusion, short of knowing how much each co-author has contributed to a publication, the $h_m$-index appears to me to be the fairest way of taking multiple authorship appropriately into account. Its determination requires a somewhat larger effort than the determination of the $h$-index, but in my opinion this is worthwhile and it could be easily incorporated into the automatic search in the Web of Science. Of course, it is essential to test the validity of the new index on the basis of more empirical data in different research fields, before it is utilized for evaluation and comparison purposes. And one should always keep in mind, that it is dangerous to measure the scientific achievements by a single number.

In summary, in my opinion the $h_m$-index is the superior way of considering multiple authorship, because in contrast to the $h_f$-index it does not need a rearrangement of the citation records, in contrast to the $h_I$-index, it is not sensitive to extreme values of the number of co-authors and cannot decrease when the number of citations increases, and in contrast to both the $h_f$- and the $h_I$-index its construction does not push highly cited papers out of the core and it enables a straightforward aggregation of data sets of several people.

The exclusion of self-citations has been shown to have a significant effect on the determination of the $h$-index (Schreiber 2007a, 2007b). It would be straightforward to exclude self-citations from the calculation of the $h_m$-index but this would necessitate a much larger effort to establish the required larger data base. The respective analysis is left for future investigations.

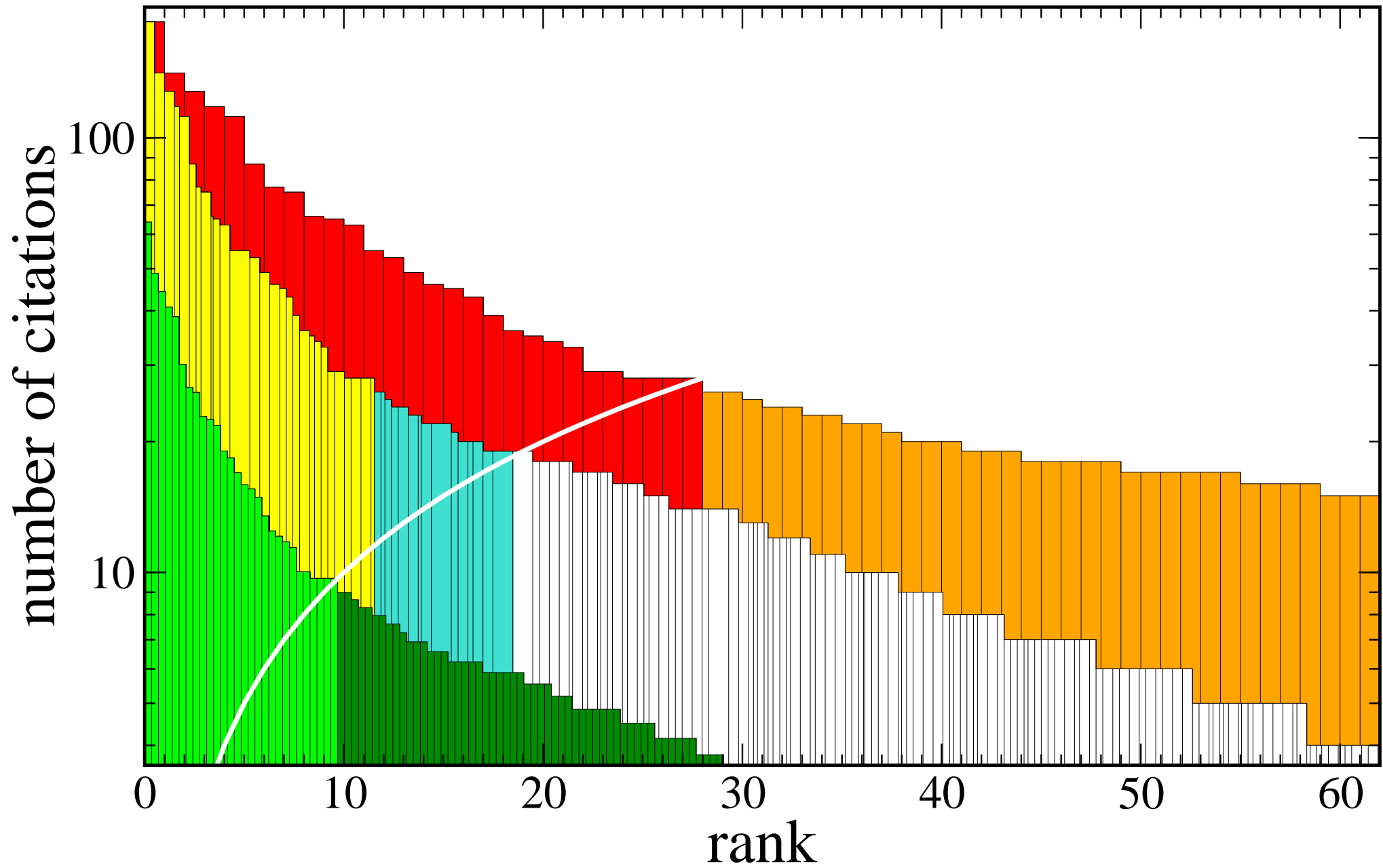

**Figure 1.** The citation counts of the most cited publications of M. Schreiber. The upper histogram with wide bars shows the numbers of citations *c(r)* versus the rank *r* which is attributed to each paper by sorting according to *c(r)*, up to the *h*-index $h = 28$ (red/dark grey) and beyond (orange/medium grey) for the 62 most-cited publications. In the middle histograms the effective rank is used so that the original histograms are compressed towards the left (yellow/light grey for the first *h* papers up to $r_{\text{eff}} = 11.53$, turquoise for the 43 papers up to the $h_m$-index $h_m = r_{\text{eff}} = 18.48$ and white beyond showing 158 papers with $c > 3$). In the lower histograms the normalization with the mean number of authors of the first *h* papers is used, so that the original histograms are compressed to the left as well as downwards (light green/medium grey up to the $h_I$-index $h_I = 9.69$ and dark green/dark grey beyond displaying 84 papers with $c > 10$). Note the logarithmic scale for *c(r)*. The thick white line displays the function $c(r) = r$, so that its intersections with the histograms (from top to bottom) yield the *h*-index, the $h_m$-index, and the $h_I$-index, respectively.

**Figure 2**. The citation counts of the most-cited 62 publications of M. Schreiber. The upper histogram is the same as in figure 1. The black circles indicate the fractionalised citation counts, i.e., the number of citations *c(r)* divided by the number of authors *a(r)* for each publication. For the lower histogram these quotients have been rearranged into decreasing order (light blue/light grey up to the $h_f$-index $h_f = r = 18$ and dark blue/dark grey beyond). The thick white line displays the function $c(r) = r$, so that its intersections with the histograms (from top to bottom) yield the *h*-index and the $h_f$-index, respectively.

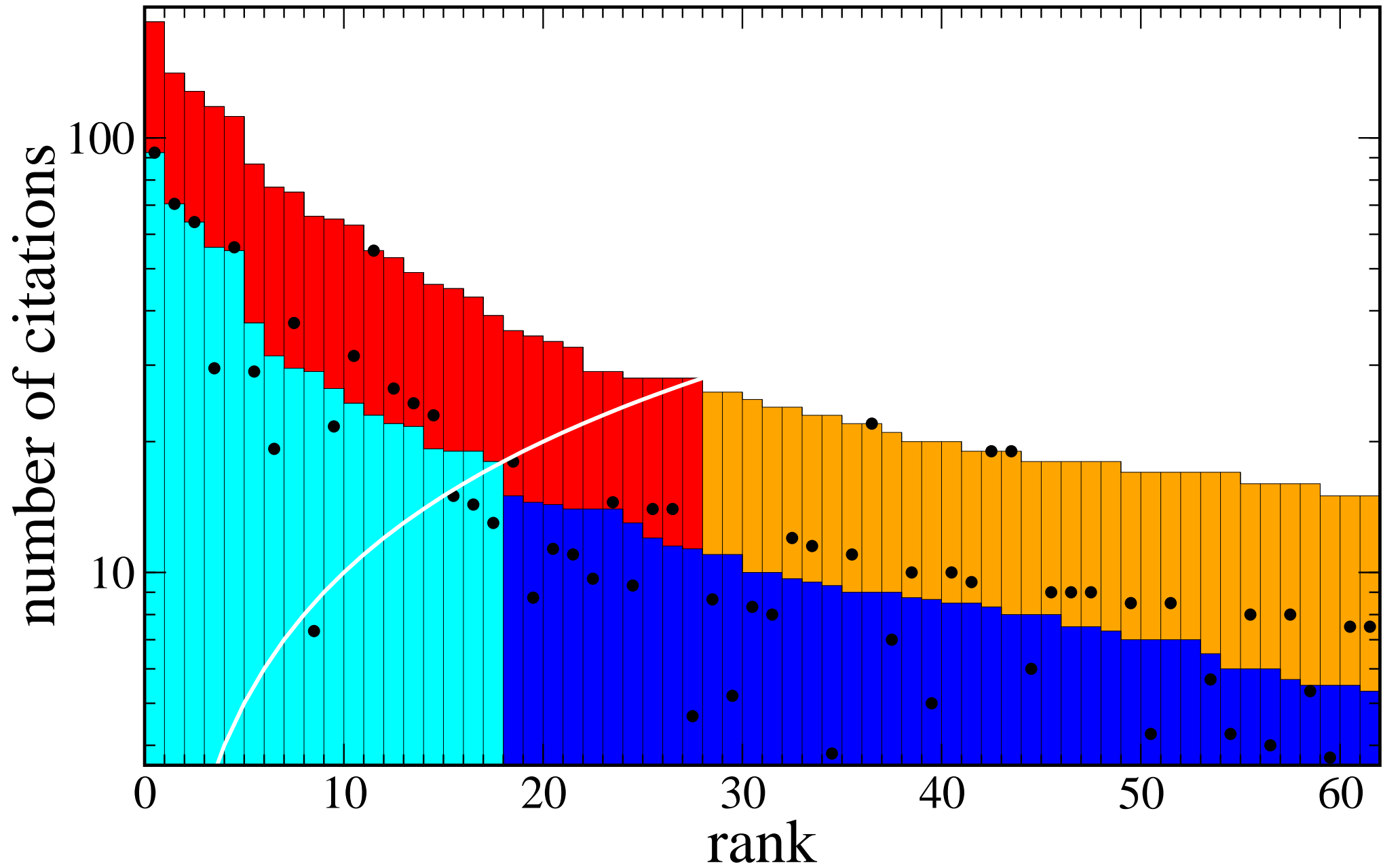

**Table 1.** A fictitious example of a model data set with 8 publications, put into order according to the number of citations $c(r)$ for rank $r$. The number of authors $a(r)$ determines the quotient $c(r)/a(r)$ which is utilized for the calculation of the $h_f$-index. The effective rank $r_{\text{eff}}(r) = \sum_{r'=1}^{r} 1/a(r')$ yields the $h_m$-index. For $r = 1$, one has $r_{\text{eff}}(1) = 1/a(1)$. For $r > 1$, the definition is equivalent to $r_{\text{eff}}(r) = r_{\text{eff}}(r-1)+1/a(r)$. Those papers which contribute to the $h$-, the $h_f$-, and the $h_m$-index are indicated by bold face in the second, fourth, and fifth column, respectively.

| $r$ | $c$ | $a$ | $c/a$ | $r_{\text{eff}}$ |
|---|---|---|---|---|
| 1 | **16** | 2 | **8** | **0.50** |
| 2 | **15** | 2 | **7.5** | **1.00** |
| 3 | **14** | 3 | **4.67** | **1.33** |
| 4 | **12** | 3 | **4** | **1.67** |
| 5 | **10** | 2 | **5** | **2.17** |
| 6 | 3 | 2 | 1.5 | **2.67** |
| 7 | 3 | 3 | 1 | **3.00** |
| 8 | 2 | 1 | 2 | 4.00 |

**Table 2.** Same as table 1, but for another data set.

| $r$ | $c$ | $a$ | $c/a$ | $r_{\text{eff}}$ |
|---|---|---|---|---|
| 1 | **9** | 3 | 3 | **0.33** |
| 2 | **9** | 3 | 3 | **0.67** |
| 3 | **9** | 3 | 3 | **1** |
| 4 | **8** | 2 | 4 | **1.5** |
| 5 | **8** | 2 | 4 | **2** |
| 6 | **8** | 2 | 4 | **2.5** |
| 7 | **8** | 2 | 4 | **3** |
| 8 | **7** | 1 | **7** | **4** |
| 9 | **7** | 1 | **7** | **5** |
| 10 | **7** | 1 | **7** | **6** |
| 11 | **7** | 1 | **7** | **7** |

**Table 3.** Same as table 1, but for another data set.

| $r$ | $c$ | $a$ | $c/a$ | $r_{eff}$ |
|---|---|---|---|---|
| 1 | **16** | 2 | 8 | **0.5** |
| 2 | **16** | 2 | 8 | **1** |
| 3 | **16** | 2 | 8 | **1.5** |
| 4 | **16** | 2 | 8 | **2** |
| 5 | **8** | 1 | 8 | 3 |
| 6 | **8** | 1 | 8 | 4 |
| 7 | **8** | 1 | 8 | 5 |
| 8 | **8** | 1 | 8 | 6 |
| 9 | 6 | 5 | 1.2 | 6.2 |

**Table 4.** Rank $r$ and effective rank $r_{eff}$ of the last paper in table 3 when its citation count $c$ is changed, and the resulting values of the indices. The values of the indices are printed in bold face if the paper contributes to the respective index.

| $c$ | $r$ | $r_{eff}$ | $h$ | $h_I$ | $h_f$ | $h_m$ |
|---|---|---|---|---|---|---|
| 6 | 9 | 6.2 | 8 | 5.33 | 8 | 6 |
| 7 | 9 | 6.2 | 8 | 5.33 | 8 | **6.2** |
| 8 | 9 | 6.2 | 8 | 5.33 | 8 | **6.2** |
| 8 | 5 | 2.2 | **8** | 4 | 8 | **6.2** |
| 9 | 5 | 2.2 | **8** | 4 | 8 | **6.2** |
| 40 | 1 | 0.2 | **8** | 4 | 8 | **6.2** |